\documentclass[aps,prl,reprint,groupedaddress]{revtex4-1}
\usepackage{amssymb,amsfonts,amsmath}
\usepackage[dvips]{graphicx}

\begin{document}


\title{Analytical solution of stochastic model of risk-spreading with global coupling}


\author{Satoru Morita}
\email[]{morita@sys.eng.shizuoka.ac.jp}
\affiliation{Department of Mathematical and Systems Engineering, Shizuoka University, Hamamatsu, 432-8561, Japan}
\author{Jin Yoshimura}
\email[]{jin@sys.eng.shizuoka.ac.jp}
\affiliation{Department of Mathematical and Systems Engineering, Shizuoka University, Hamamatsu, 432-8561, Japan}
\affiliation{Marine Biosystems Research Center, Chiba University, Uchiura, kamogawa, Chiba 299-5502, Japan}


\date{\today}

\begin{abstract}
We study a stochastic matrix model to understand the mechanics of 
risk-spreading (or bet-hedging) by dispersion. 
Such model has been mostly dealt numerically
except for well-mixed case, so far.
Here, we present an analytical result, 
which shows that optimal dispersion leads to Zipf's law. 
Moreover, we found that the arithmetic ensemble
average of the total growth rate converges to the geometric one,
because the sample size is finite.
\end{abstract}

\pacs{87.10.Mn, 87.23.Cc, 89.65.Gh}

\maketitle

\subsection{INTRODUCTION}

Environmental heterogeneity and fluctuations can induce nontrivial effects 
in ecological \cite{r1,r2,r3,r4,r5,r6,r7,r8} and economic 
systems \cite{r9,r10,r11}. Risk-spreading (or
bet-hedging) is one of the most important concepts about
strategy in unpredictably fluctuating environments. 
For example, consider an ecological model of 
offspring allocation into two habitats, either of which is so poor that 
the population cannot survive if they are exclusive \cite{r4,r8}. 
In this model, 
a dispersal can lead to population persistence. In economic/financial 
situations, diversification is a key component of long-term investment 
strategies. To minimize a risk and increase expected returns, the 
wealth has to be invested into various types of funds or asset classes. 
The purpose of this paper is 
to present theoretical results about the long-term growth 
for complicated systems and to make clear the mechanism of risk-spreading.

In this paper, we study a discrete-time stochastic matrix model for 
populations that inhabit in $n$ discrete habitats \cite{r7}. This class of 
models have been dealt mostly numerically by simulations. Recently, 
the authors proposed an analytical theory for a population that inhabits 
in two habitats \cite{r8}. In this case, the ratios of populations among 
habitats are distributed in a complicated self-similar manner, while 
the marginal distribution of each population is a lognormal distribution. 
Here, we expand the previous theory for the system with a lot 
of habitats. We show that populations in habitats at the same 
time follows a power law distribution, though the ensemble distribution of 
population at each habitat converges to a lognormal distribution in 
the limit of long time. 
Especially, when the dispersion rate is optimal, 
the distribution of population tends to follow the famous Zipf's law, where 
the exponent of the power law tail is near one. 
Here, to characterize the fitness or return of 
the entire system, we use two different indices: the geometric mean 
and the arithmetic one of the total growth rate. The inequality 
of arithmetic and geometric average leads that the arithmetic one 
is greater than or equal to the geometric one. We study analytically 
and numerically the characteristics of these two indices in detail.
As a result, we show these two indices coincide by averaging 
over adequately long time.

\section{Model}
Let us consider population that inhabits in $n$ discrete habitats. 
Let $x_i (t)$ be the number of individuals in habitat $i \ (1\leq i\leq n)$
 at time $t$. Thus, the state of the population is described by a 
vector $(x_1 (t),x_2 (t),\dots,x_n (t))$. 
In each habitat, the population reproduces with random growth 
rates $m_i (t)$. 
The habitats are connected to one another by corridors, which are 
described by the adjacency matrix $a_{ij}$. 
A fraction of the population disperses from a habitat $j$ to 
another habitat $i$. Thus, the population dynamics is given as 
\begin{equation}
 x_i(t+1)=(1-q)m_i(t)x_i(t)+qs\sum_{j=1}^n l_{ij} m_j(t)x_j(t),
\label{e01}
\end{equation}
where $l_{ij}$ is the link matrix defined by using the adjacency matrix:
\begin{equation}
l_{ij}=\frac{a_{ij}}{\sum_{k=1}^n a_{kj}}.
\label{e02}
\end{equation}
The parameters $q$ and $s$ are between 0 and 1. 
The migration rate $q$ represents the proportion of population that 
migrates from a habitat to other habitats. When $q=0$, all the habitats 
are isolated completely. The parameter $s$ is the survival rate during 
migration between habitats. 
Here, the local growth rates $m_i (t)$ are stochastic variables with 
finite variance. The probability distributions of $m_i (t)$ are the 
same for all habitats. We assume that the local growth rates $m_i (t)$
 have no temporal and spatial correlation for simplicity. 
If this model is interpreted as a model of financial economics, 
$x_i (t)$ represents money that is invested into enterprise $i$. 
In this case, the migration rate $q$ and the link matrix $l_{ij}$ 
give a portfolio strategy and $(1-s)$ can be regarded as a kind 
of transaction costs.

In this paper, we focus on the case of global coupling
\begin{equation}
a_{ij}=\left\{\begin{array}{ll}
1 & (i\neq j)\\
0 & (i=j).
\end{array} \right.
\label{e03}
\end{equation}
In this case, it is convenience to consider the mean field
\begin{equation}
h(t)=\frac{1}{n}\sum_{i=1}^n x_i(t).
\label{e04}
\end{equation}
Summing \eqref{e01} over all habitats, we obtain the dynamics of the mean field
\begin{equation}
h(t+1)=(1-q+sq)\bar{m}(t)h(t).
\label{e05}
\end{equation}
Here $\bar{m}(t)$ is the whole growth at time $t$, which is defined as
\begin{equation}
\bar{m}(t)=\frac{1}{n}\sum_{i=1}^{n}m_i(t)y_i(t),
\label{e06}
\end{equation}
where $y_i(t)$ is the relative population size
\begin{equation}
y_i(t)=\frac{x_i(t)}{h(t)}.
\label{e07}
\end{equation}
Neglecting small terms of $O(n^{-1})$, 
the dynamics of the relative population size is given by
\begin{equation}
y_i(t+1)=\frac{m_i(t)(1-q)}{\bar{m}(t)(1-q+sq)}y_i(t)+sq\frac{1}{1-q+sq}.
\label{e08}
\end{equation}
Consequently, the dynamics of the population is described by \eqref{e05},
 \eqref{e06} and \eqref{e08}. 
Taking into consideration \eqref{e04} and \eqref{e07}, 
we have the following restriction condition
\begin{equation}
\frac{1}{n}\sum_{i=1}^{n}y_i(t)=1.
\label{e09}
\end{equation}
The initial condition is set as $x_i(0)=1$, 
i.e., $h(0)=1$.

\section{Results}
\subsection{Approximation}
To solve the dynamics of the population, 
we assume that $n$ is adequately large, such that we can neglect 
the dependence of $\bar{m}(t)$ on each $y_i (t)$ at the same time $t$. 
In this approximation, eq.~\eqref{e08} can be treated as a 
one-dimensional stochastic process.
Thus, the distribution of each relative population size $y_i$ 
converges to a stationary distribution. 
Eq.~\eqref{e08} is regarded as 
a multiplicative process with an additional constant term.
If the probability that the multiplicative factor 
$m_i (t)(1-q)/[\bar{m}(t)(1-q+sq)]$ is larger than one is positive, 
the stationary distribution has a power law tail \cite{add0,r12,r13,r14,r15}
\begin{equation}
p(y)\sim y^{-\alpha-1}.
\label{e10}
\end{equation}
The distribution with a power law is said to follow Pareto distribution \cite{r16}. 
In the case of $\alpha\leq 2$, the variance of $y_i (t)$
diverges in the limit of $n\to\infty$. 
Even in this case, the variance of $\bar{m}(t)$ remains finite. 
Indeed, from the restriction \eqref{e09}, 
it is obvious that the variance of $\bar{m}(t)$ is never larger than 
one of $m_i(t)$. 
From the generalized central limit theorem \cite{r17}, 
we obtain that for large $n$, the variance of $\bar{m}(t)$ obeys
\begin{equation}
V(\bar{m}(t))\sim n^{-\mu},
\label{e11}
\end{equation}
where
\begin{equation}
\mu=\left\{\begin{array}{ll}
\alpha-1 & (1<\alpha<2)\\ 
1 & (\alpha \geq 2).
\end{array}\right.
\label{e12}
\end{equation}

Taking the logarithms of both sides of eq.~\eqref{e05} and summing 
them from $t=0$ to $t=T-1$, we obtain
\begin{equation}
\log h(T)=\log h(0)+T \log(1-q+sq)+\sum_{t=0}^{T-1}\log\bar{m}(t).
\label{e13}
\end{equation}
Thus, $\log h(T)$ is determined by using the sum of the whole growth rate 
$\log\bar{m}(t)$. 
Since $\log\bar{m}(t)$ is treated as a time series of random variables, 
the ensemble distribution of $\log h(T)$ is expected to converge to a 
normal distribution. 
In other words, $h(T)$ follows a lognormal distribution for large $T$. 
A way to characterize the fitness or return of the system is to 
calculate the geometric average of the total growth rate as
\begin{equation}
\frac{1}{T} E(\log h(T))=\log(1-q+sq)+E(\log\bar{m}(t)),
\label{e14}
\end{equation}
where $E$ denotes the ensemble average.
The reason for this terminology is that \eqref{e14} is the logarithm 
of the geometric ensemble average of $h(T)$. 
Since $\bar{m}(t)$ has ergodicity, the ensemble average 
$E(\log\bar{m}(t))$ is the same as its time average. 
Note that $E(\log h(T))$
coincides with the logarithm of the median of $h(T)$,
because $h(T)$ follows a lognormal distribution.
Moreover, 
the second term in the right hand side of \eqref{e14} is the logarithm of 
the geometric average of $\bar{m}(t)$. 
For adequately large value of $T$, the ensemble variance of the whole 
growth rate is given by
\begin{equation}
\frac{1}{T} V(\log h(T))=V(\log\bar{m}(t)).
\label{e15}
\end{equation}
Because of ergodicity, the ensemble variance 
$V(\log\bar{m}(t))$ is also the same as its time variance.
For a random variable $X$ following a lognormal 
distribution, we have a formula
\begin{equation}
\log(E(X))=E(\log X )+\frac{1}{2} V(\log X).
\label{e16}
\end{equation}
By using eqs.~\eqref{e15} and \eqref{e16},
we obtain
the following relation
\begin{equation}
\frac{1}{T} \log E(h(T))=\frac{1}{T} E(\log h(T))+\frac{1}{2} 
V(\log \bar{m}(t)).
\label{e17}
\end{equation}
Another way to characterize the fitness or return of the system is to 
calculate eq.~\eqref{e17}, that is the arithmetic average of the 
total growth rate. 
Because $V(\bar{m}(t))$ follows \eqref{e11} in the large size limit 
($n\to\infty$) from \eqref{e11}, $V(\log\bar{m}(t))\sim n^{-\mu}$.
As a result, for $n\to\infty$, the arithmetic ensemble average \eqref{e17}
is expected to converge to its geometric ensemble average \eqref{e14}. 

In the case that  the autocorrelation of $\bar{m}(t)$ is negligibly weak, 
i.e., $\bar{m}(t)$ and $\bar{m}(t')$ are independent of each other 
for $t\neq t'$, eq.~\eqref{e05} leads 
\begin{equation}
\frac{1}{T} \log E(h(T))=\log(1-q+sq)+\log E(\bar{m}(t)).
\label{e18}
\end{equation}
We call eq.~\eqref{e18} well-mixed approximation, 
because it is exactly correct for well-mixed system (i.e., $1-q=sq/(n-1$)) \cite{r7}.

\subsection{Numerical Simulations}
To examine the above theoretical predictions, we perform numerical 
simulations. 
For simplicity, the local growth rate $m_i (t)$ ($i=1,2\dots,n$) 
takes one of two values, $m_-$ with probability $p$ and $m_+$ with 
probability $1-p$ independently, where we set $m_-<m_+$. 
From \eqref{e06}, 
the average $E(\bar{m}(t))$ is simplified to
\begin{equation}
E(\bar{m}(t))=p m_- + (1-p) m_+.
\label{e19}
\end{equation}
On the other hand, $E(\log\bar{m}(t))$ does not have an analytical form.
\begin{figure}
\begin{center}
\includegraphics[width=0.4\textwidth]{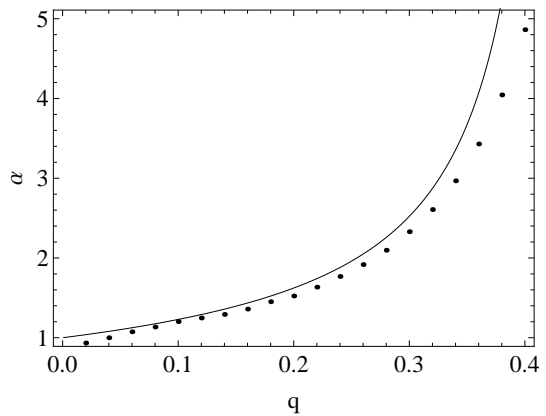}\\%
\end{center}
\caption{ 
The exponent $\alpha$ of the power law tail as a function of $q$. 
Here, we estimate $\alpha$ by using the top 0.1\% 
for $n=10^5$ and 20 samples \cite{r16}.
The other parameters are set as $m_+=2$, $m_-=0.1$, $p=0.3$, $s=0.5$. 
The circles represent the numerical estimated values and the curve 
gives the theoretical relation from \eqref{e21}.
\label{f1}}
\end{figure}
%
Figure \ref{f1} shows the exponent $\alpha$ of the power law. 
The condition to realize steady power law distribution leads \cite{r13}
\begin{equation}
E\left(\left(\frac{m_i(t)(1-q)}{\bar{m}(t)(1-q+sq)}\right)^{\alpha} \right)=1.
\label{e20}
\end{equation}
If $V(\bar{m}(t))$ is relatively small compared to $E(\bar{m}(t))$, 
eq.~\eqref{e20} is approximated by
\begin{equation}
\frac{[{m_-}^{\alpha} p+{m_+}^{\alpha}(1-p)](1-q)^{\alpha}}
{[(pm_- +(1-p) m_+)(1-q+sq)]^{\alpha}}=1.
\label{e21}
\end{equation}
Thus, we can estimate the value of $\alpha$ and we obtain the 
approximation of $\mu$ by using \eqref{e12}. 
Equation \eqref{e21} indicates that 
the exponent $\alpha$ 
converges to 1 in the limit $q\to 0$.
Note that in the case
\begin{equation}
\frac{m_+ (1-q)}{(pm_-+(1-p) m_+)(1-q+sq)}<1,
\label{e22}
\end{equation}
\eqref{e21} has no relevant solution, and   
the tail of the stationary distribution is not power law, 
but decays fast. In this case, $\mu=1$. 
The curve in Fig.~\ref{f1} represents the theoretical prediction \eqref{e21}, 
which fits the numerical results quite nicely.

\begin{figure*}
\begin{center}
\includegraphics[width=.98\textwidth]{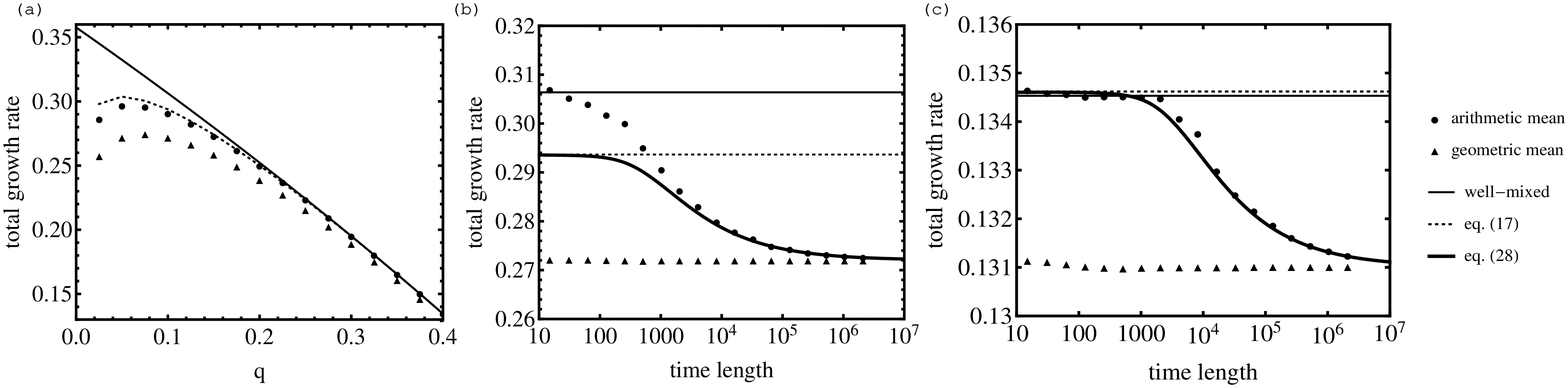}\\%
\end{center}
\caption{(a) Indices to measure the total growth rate 
as a function of $q$ for $n=100$ and
$T=2^{10}$ after a transient of $100,000$ time steps. 
(b), (c) These indices as a function of time interval $T$ for $q=0.1$
(b) and q=0.4 (c). 
The other parameters are same as Fig.~\ref{f1}. 
The circles and triangles represent the arithmetic and 
the geometric ensemble average of the growth rate, respectively. 
Here, the ensemble consists of $N_s=10^5$ samples. 
The plots are given by the average over 10 such realizations. 
The dotted lines are the prediction (17) for the arithmetic average, 
where $E(\log\bar{m}(t))$ and $V(\log\bar{m}(t))$ are calculated over
the time interval 0 to $10^7$.
The thin lines are the well-mixed approximation \eqref{e18}.
The dotted line is the prediction \eqref{e17}, 
where $E(\log\bar{m}(t))$ and $V(\log\bar{m}(t))$ are calculated over 
the time interval 0 to $10^7$. 
The thick curves in (b) and (c) represent the result \eqref{e28}
taking into account the deviation due to finite sample size.\label{f2}}
\end{figure*}

Figure \ref{f2} shows the arithmetic and the geometric ensemble averages of 
the total growth rate. In Fig.~\ref{f2} (a), both of them have a peak
(the circles and triangles represent the arithmetic and the 
geometric averages). 
This means that the dispersion rate has a finite optimal value. 
The solid curve in Fig.~\ref{f2} (a) corresponds to the well-mixed 
approximation \eqref{e18}, 
which slopes downward to the right when the survival rate $s$ 
during migration between habitats is smaller than one.
When $n$ increases, both the arithmetic and the geometric averages 
approach to the curve of the well-mixed approximation (not shown). 
Consequently, when the system is large ($n\gg1 $)
and the survival rate $s$ is low, 
the optimal value of the dispersion rate is near 0, i.e., 
the exponents $\alpha$ are close to 1 (see Fig. \ref{f1}).

From Fig.~\ref{f2} (a), the arithmetic average seems to agree with the 
prediction \eqref{e17} (dotted curve). 
However, when the time length $T$ is larger, 
the arithmetic average deviates from \eqref{e17} and approaches to 
the geometric average (Fig.~\ref{f2} (b) and (c)). 
We can interpret this result as the effect of finite-size sampling. 
In Fig.~\ref{f2}, the sample size $N_s$ is set to $10^5$ in calculating the 
ensemble average.
 
Recalling that if a stochastic variable $x$ follows the standard normal 
distribution, the maximum value $x_{\max}$ in $N_s$ 
samples is estimated approximately by
\begin{equation}
\frac{1}{\sqrt{2\pi}}\int_{x_{\max}}^{\infty}
\exp\left(\frac{-x^2}{2}\right)dx \simeq \frac{1}{N_s}.
\label{e23}
\end{equation}
For $N_s=10^5$, we calculate $x_{\max}\sim 3.719$. 
The stochastic variable $(\log h(T)- E(\log h(T)))/\sqrt{V(\log h(T))}$ 
follows the standard normal distribution. 
Thus, the maximum value of ($\log h(T)-E(\log h(T)))/\sqrt{V(\log h(T))}$ 
is also $x_{\max}$. Using eq.~\eqref{e15}, 
the maximum value of $\log h(T)$ is estimated as 
\begin{equation}
\frac{1}{T} \max(\log h(T))\simeq \frac{1}{T} E(\log h(T))+
x_{\max} \sqrt{V(\log\bar{m}(t))/T}.
\label{e24}
\end{equation}
If $V(\log\bar{m}(t))$ is larger than $2x_{\max}\sqrt{V(\log\bar{m}(t))/T}$,
then the average \eqref{e17} is over the maximum \eqref{e24}. 
However, this is impossible. 
This fact indicates that the prediction \eqref{e17} gives a wrong result, 
unless $T$ is smaller than a critical time
\begin{equation}
T_c\simeq\frac{4{x_{\max}}^2}{V(\log\bar{m}(t))}.
\label{e25}
\end{equation}
Since we can estimate $x_{\max}\sim\sqrt{\log N_s}$ for large ensemble 
$N_s$ \cite{r18}
and $V(\log\bar{m}(t))\simeq V(\bar{m}(t))\sim n^{-\mu}$ (see (11)), 
the critical time $T_c$ has a scaling relation
\begin{equation}
T_c \sim n^{\mu} \log N_s.
\label{e26}
\end{equation}
When $T>T_c$, the average $E(h(T))$ can be described by an integral 
with the interval from $-\infty$ to $x_{\max}$
\begin{equation}\begin{array}{ccl}
E(h(T)) & \simeq & \displaystyle \frac{1}{\sqrt{2\pi}}
\int_{-\infty}^{x_{\max}}\exp\left(\frac{-x^2}{2}\right)\\
& & \exp\left(E(\log h(T))+ x\sqrt{V(\log h(T))}\right) dx.
\end{array}
\label{e27}
\end{equation}
By a simple algebra, we obtain
\begin{equation}\begin{array}{l}
\displaystyle \frac{1}{T} \log E(h(T)) \simeq  \displaystyle
\frac{1}{T} E(\log h(T))+\frac{1}{2} V(\log\bar{m}(t))\\
  \displaystyle +\frac{1}{T} \log\left[\frac{1}{\sqrt{2\pi}} 
\int_{-\infty}^{x_{\max}-\sqrt{T V(\log\bar{m}(t))}}
\exp\left(\frac{-x^2}{2}\right) dx\right].
\end{array}
\label{e28}
\end{equation}
In Fig.~\ref{f2} (b) and (c), thick curves represent the estimation \eqref{e28}. 
The theoretical result agrees with the numerical simulation when $T$ is large.
On the other hand, for small $T$, a deviation is observed. 
In this case, the distribution of $\log h(T)$ does not yet converge to 
the normal distribution, thus \eqref{e28} cannot give a good approximation, 
especially when the dispersion rate $q$ is small (Fig.~\ref{f2} (b)). 
In the limit $T\to\infty$, the third term in the left hand size in \eqref{e28}
converges to $-1/2 V(\log\bar{m}(t))$. Thus, $\log E(h(T))$ is asymptotic to
$E(\log h(T))$ for large $T$. 
Consequently, after a long time ($T\gg T_c$), 
the arithmetic average of the growth rate coincides with the geometric one
essentially. 

\section{Discussions}
Up to here, we have given an analytical solution for a discrete-time 
stochastic matrix model of risk-spreading and compare it with numerical 
simulations. 
The population in each habitat follows a lognormal distribution. 
In high-dimensional systems ($n\gg 1$), 
the simultaneous distribution of populations at the same time has 
a power law tail. 
Since the population follow a power law distribution, 
the network resulting from the migration among the habitats
has scale free structure \cite{a02,a03}.
In the case that the survival rate $s$ during migration 
between habitats is lower than one, 
the optimal dispersion rate is near 0 for large $n$. 
Equation \eqref{e21} indicates that 
when $q$ approaches to 0, the exponent $\alpha$ 
of the power law approaches one. 
The power law distribution with $\alpha\simeq 1$ 
is known as Zipf's law \cite{r19},
which has been found in various fields, including the population 
of cities \cite{r20,a01} and the asset of companies \cite{r21}. 
Earlier studies showed that for the random growth model without migration, 
Zipf's law is realized 
when the mean growth rate is close to zero \cite{r20,a01}.
Our result shows that taking into account the migration with cost,
Zipf's law is given as the optimal solution of the risk-spreading model. 
We therefore conclude that natural selection,
which will favor the trait with high fitness, can lead to 
Zipf's law when there is cost during the migration.
As far as we know, this is the first report that links Zipf's law with
risk-spreading adaptation.

To characterize the fitness or return, 
we used two types of indices: the geometric ensemble mean and 
the arithmetic one of the total growth rate.
In evolutionary ecology, 
natural selection is considered as optimizing fitness.
As fitness measure, the arithmetic mean has been used traditionally. 
On the other hand, the geometric mean has been used 
as the fitness measure across a number of generations \cite{r23,r24}.
In our model, the difference between values of these two indices is
in the order of $n^{-\mu}$.
We found that the geometric mean hardly depends on the 
time length over which is averaged,
though the arithmetic mean is near the well-mixed approximation 
when the time length is very short,  
and converges to the geometric mean in the long-time limit.
This result demonstrates the validity of using the geometric mean
to estimate the long-term fitness.
The boundary between short and long terms, 
which is given by eq.~\eqref{e26}, is proportional to $n^{\mu}$.
This suggest that when the difference between the arithmetic and 
geometric mean is larger for short time length, the convergence tends
to be faster.

We studied the case of global coupling to be tractable analytically.
We also perform some numerical simulations for other networks: 
regular random graphs and lattice. 
According to the numerical results, the results observed in Fig.~\ref{f2} 
do not change qualitatively with the network structures. 
In all cases, the arithmetic average of the total growth rate 
converges to the geometric one for an adequately long time length, 
because of finite sample size. When the degree of coupling decreases, 
the optimal migration rate increases. 
The restriction on the coupling suppresses power law. 
Especially in the case of lattice, eq.~\eqref{e11} 
does not hold and the decrease of $V(\bar{m}(t))$ 
with $n$ stops around a critical size.
It remains future work to analyze the model \eqref{e01} 
for complex networks including heterogeneous networks such as 
scale-free networks.

\begin{acknowledgments}
This work was supported by grants-in-aid from the Ministry of Education, 
Culture, Sports, Science and Technology of Japan to S.M. (No. 24500273) 
and J. Y. (No. 22370010 and No. 22255004). A part of the numerical 
computation in this work was carried out at the Yukawa Institute 
Computer Facility.
\end{acknowledgments}

\bibliography{ref.bib}

\end{document}